\documentclass{mrf}

\usepackage{graphics} 
\usepackage{amsmath}
\usepackage{subfig}
\usepackage{epstopdf}

\begin{document}

\supertitle{research paper}

\title[Radar tracking for granular particles]{Correction of IQ mismatch for a particle tracking radar}

\author[Granular particle tracking with radar] {Felix Rech$^{1}$ and Kai Huang $^{1,2}$}

\address{\add{1}{Experimentalphysik V, Universit\"at Bayreuth, 95440 Bayreuth, Germany}
\add{2}{Division of Natural and Applied Sciences, Duke Kunshan University, No. 8 Duke Avenue, Kunshan, Jiangsu, China 215316}}

\corres{\name{Kai Huang}
\email{kai.huang@uni-bayreuth.de}}

\begin{abstract}
For a better understanding of granular flow problems such as silo blockage, avalanche triggering, mixing and segregation, it is essential to have a `microscopic' view of individual particles. In order to cope with the difficulty arising from the opacity of granular materials, such as sands, powders and grains, a small scale bi-static radar system operating at $10$\,GHz (X-band) was recently introduced to trace a sub-centimeter particle in three dimensions. Similar to a moving target indicator radar, the relative movement of the tracer with respect to each of the three receiving antennae is obtained via comparing the phase shift of the electromagnetic wave traveling through the target area with an IQ-Mixer. From the azimuth and tilt angles of the receiving antennae obtained in the calibration, the target trajectory in a three-dimensional Cartesian system is reconstructed. Using a free-falling sphere as a test case, we discuss the accuracy of this radar system and possible ways to enhance it by IQ mismatch corrections.
\end{abstract}

\keywords{ Authors should not add keywords, as these will be chosen during the submission process (see http://journals.cambridge.org/\-data/\-relatedlink/\-MRF\_topics.pdf for the full list)}

\maketitle

\section{Introduction}
Since the beginning of last century, radar technology has been continuously developed and benefiting us in many different ways: From large scale surveillance radar systems that are crucial for aircraft safety and space exploration \cite{Skolnik2001}, to small scale systems for monitoring insects \cite{Oneal2004}. Considering the limit of radar tracking technique, it is intuitive to ask: How small can an object be accurately tracked by a radar system? Can it be as small as a tracer particle with a size comparable to a grain of sand? If this challenge can be solved to a satisfactory level, radar technology can help us greatly in a better monitoring, understanding, and predicting granular flow problems that widely exist in nature, industry and our daily lives, ranging from geo-science (e.g. landslide, debris flow), through chemical and civil engineering (e.g. pile drilling), to space exploration (e.g. landing on an asteroid) \cite{Duran2000,pg13}. The reason behind is that the mobility of single particles in a granular material can influence dramatically its collective behavior, owing to its discrete nature as well as heterogeneous distributions of force chains inside \cite{Bi2011}. 

In the past decades, there have been substantial progresses in imaging granular particles \cite{Amon2017}. Due to the fact that most of the granular materials are opaque, optical means for imaging particles in three dimensions (3D), such as refractive index matching \cite{Dijksman2017}, are very limited. Instead, X-ray tomography \cite{Athanassiadis2014} and Magnetic Resonance Imaging \cite{Stannarius2017} have both been used to identify the internal structures of granular materials. However, the limited time resolution of scanning technique as well as the huge amount of data to be processed hinder the investigation of granular dynamics. For the investigation of granular dynamics, alternative approaches including radar systems for tracer particles have also been discussed \cite{Hill2010}.

Recently, we introduced a small scale continuous-wave (CW) radar system working at X-band to track a spherical object with a size down to $5$\,mm \cite{Ott2017}. In comparison to other techniques, the continuous trajectory of a tracer particle obtained by the radar system helps in deciphering granular dynamics greatly. Here, we characterize the uncertainty of this system, discuss possible sources of error and ways to improve the accuracy.

\section{Experimental set-up and tracking algorithm}
\begin{figure}[!ht]

\centering
\includegraphics[width=0.85\columnwidth]{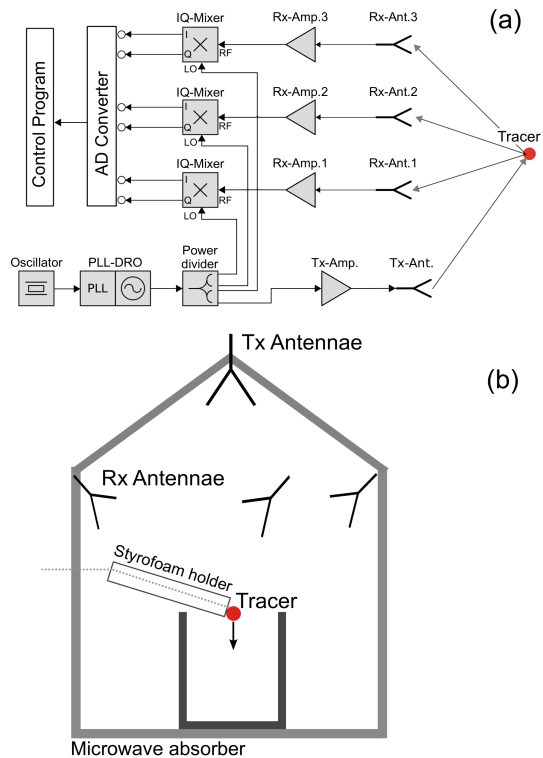}
\caption{Block diagram of the radar system (a) and a sketch showing the configuration of the transmission (Tx.) and receiving (Rx.) antennae as well as the tracer holding device.}
\label{fig:setup}
\end{figure}

Figure~\ref{fig:setup} shows the block diagram of the radar system (a) together with a sketch of the set-up used to test the accuracy of reconstructed trajectories (b). The bi-static radar system operates at 10\,GHz (X-band) with one transmission (Tx.) antenna pointing in the direction of gravity (defined as $-z$ direction) and three receiving (Rx.) antennae mounted symmetrically around the $z$ axis. Polarized electromagnetic (EM) waves, after being scattered by the tracer particle, are captured by the Rx. antennae. With the help of an IQ mixer, which compares the phase shift between the Tx. and Rx. antennae, the change of the absolute traveling distance for the $i$th antenna $L_i = l_0 + l_i$ can be obtained, where $l_0$ and $l_i$ are the distance between Tx. Ant. and the target and the distance from the target to the Rx. Ant, respectively. Subsequently, a transformation matrix $\vec T$ is applied to the distance vector $\vec L = (L_1,L_2,L_3)$ to reconstruct the tracer trajectory in a 3D Cartesian system. The smallest spherical object being traceable by the system was found to be $\sim 5$\,mm, in agreement with the prediction of radar equation. More detailed descriptions of the radar system can be found in \cite{Ott2017}.

IQ-Mixers play an essential role in the accurate ranging of a target. The LO and RF inputs correspond to the signal sent to the Tx. antenna [$a\cos(2\pi f_{\rm 0}t)$] and that received by a Rx. antenna [$b\cos(2\pi ft + \theta)$], where $a$ and $b$ are the magnitudes of the corresponding signals, $f_{\rm 0}$ and $f$ are the transmitted and received signal frequencies. The output signals of the IQ-Mixer are

\begin{eqnarray}
I=\frac{ab}{2}\cos[2\pi(f_{\rm 0}-f)t - \theta], \nonumber \\
Q=\frac{ab}{2}\sin[2\pi(f_{\rm 0}-f)t - \theta]. \label{eq:iq2}
\end{eqnarray}

\noindent Subsequently, the relative movement of the tracer is obtained from the phase shift of $I+Qi$ in a complex plane. If $L_i$ varies with a distance of one wavelength, the vector $I+Qi$ rotates $2\pi$. As IQ mixers provide analogue signals that representing the mobility of the tracer, the time resolution of the radar system is only limited by the analogue-digital (AD) converter. 

Although distance measures rely only on the phase information, the sensitivity and accuracy of the system depend on IQ signal strength. In order to have a sufficient signal to noise ratio, the directions of the horn antennae (Dorado GH-90-20) are adjusted with the help of a laser alignment and range meter (Umarex GmbH, Laserliner) to face the target area. According to the specification of the antenna, the main lobe of its radiation pattern has an opening angle of $\sim15$ degrees. Thus, we estimate the field of `view' of the radar system has a volume of about $30 {\rm cm} \times 30 {\rm cm} \times 30 {\rm cm}$, taking into account the average working distance of the antennae. The distance between each antenna and the center of the coordinate system is also measured by the laser meter during the adjustment process. The polarization of the antennae are adjusted to maximize the raw I and Q signals. The whole system is covered with microwave absorbers (Eccosorb AN-73) to reduce clutter and unwanted noises from the surrounding. In addition, the container for granular materials and the holder of the tracer are made of expanded polystyrene (EPS), which is transparent to EM waves. 

A metallic sphere with a diameter $d=$10\,mm is used as the tracer. It is initially held by a thin thread wrapped around and released by gently pulling the thread such that the initial velocity of the falling sphere is close to $0$. This design enables a defined and reproducible initial condition for a comparison among various experimental runs. The raw IQ signals from the AD converter (NI DAQPAD-6015) are recorded and further processed with a Matlab program to obtain the reconstructed trajectories. 

\section{Data analysis and error correction}

\begin{figure}
\centering
\vskip -2em
\includegraphics[width=0.85\columnwidth]{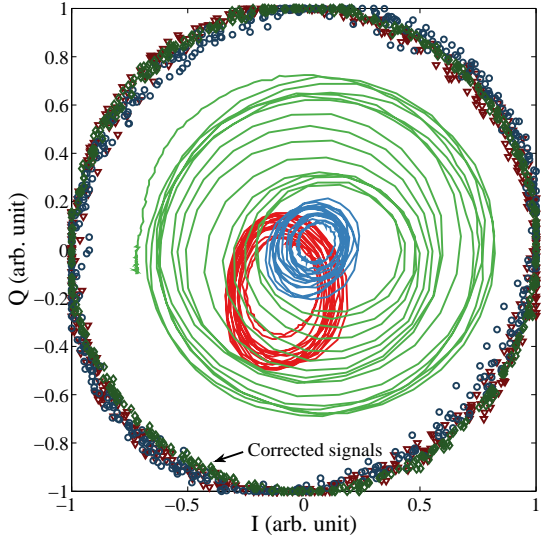}
\caption{Raw (continuous lines) and corrected (open symbols) signals representing a free-falling sphere from a height of $27$\,cm. Red (dark red), green (dark green) and blue (dark blue) curves (points) correspond to the results from channel 1, 2, and 3, respectively. For a better visibility, the offsets of the raw signals are removed. }
\label{fig:raw}
\end{figure}

As Fig.\,\ref{fig:raw} shows, the raw IQ signals are typically not ideal in the sense that the in-phase (I) and quadrature (Q) signals are not always orthogonal with each other. This mismatch may arise from the DC offsets of either I or Q signal, gain and phase imbalance. How to correct such kind of errors has been extensively discussed in, for instance \cite{Churchill1981} or \cite{Huang2002}, particularly along with the development of telecommunication and non-invasive motion detecting techniques \cite{Park2007}. The distortions are typically attributed to device imperfections as well as clutter. However, for the system being used here, there are additional errors arising from the mobility of the tracer itself, which can not be readily corrected with a calibration of the hardware. Moreover, distortion may also arise from the interaction of the scattered signal from the tracer with that from static objects that are not completely transparent to EM waves. In that case, the existence of `mirrored' particles may lead to additional uncertainty.    

\begin{figure}
\centering
\includegraphics[width=0.85\columnwidth]{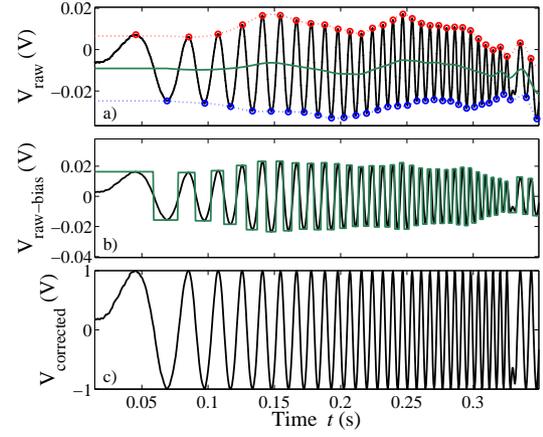}
\caption{Process for IQ mismatch correction. (a) A representative raw signal with peaks and valleys marked with red and blue open circles. From an average of both spline fits for the peaks and valleys, the bias error (green line) for the raw signal as a function of time is estimated. (b) Bias corrected signal time dependent rescaling factors (green line) for the correction of gain error. (c) Corrected signal for further analysis.}
\label{fig:corr}
\end{figure}

Here, we use the following approach to correct IQ mismatch arising from multiple sources of errors. It works best when the object moves in a distance covering multiple wavelengths. As illustrated in Fig.\,\ref{fig:corr}, the correction process is composed of the following steps: First, we identify the time segment of the raw data $V_{\rm raw}$ that contains the movement of the tracer particle via finding the start and end of the fluctuations. Second, the peaks (red circles) and valleys (blue circles) of individual fluctuations are determined by finding the local extreme values in the selected data. Third, the bias error $V_{\rm bias}$ [green line in (a)] is estimated as the mean value of the spline fits of peaks and valleys (dashed lines). In order to avoid unrealistic extrapolations, the bias error starts to vary only from the first peak. Fourth, the bias error is removed and the corrected signal $V_{\rm raw - bias}$ is segmented by zero crossings. Finally, the data in individual segments are rescaled by local maxima and minima to correct gain mismatch.   

As shown in Fig.\,\ref{fig:raw}, this approach can effectively find time dependent correction factors due to tracer movement. For the corrected data of channel 3 (dark blue circles), there exists a slight deviation from a perfect circle, indicating the existence of a small phase error. This arises presumably from the fact that perfect polarization cannot be achieved for all three Rx. antennae. Further investigations are needed to check whether this error can be avoided by using circular polarized EM waves or by correcting the phase error between I and Q signals in the Matlab program.

\begin{figure}
\vskip -3em
\centering
\includegraphics[width=0.85\columnwidth]{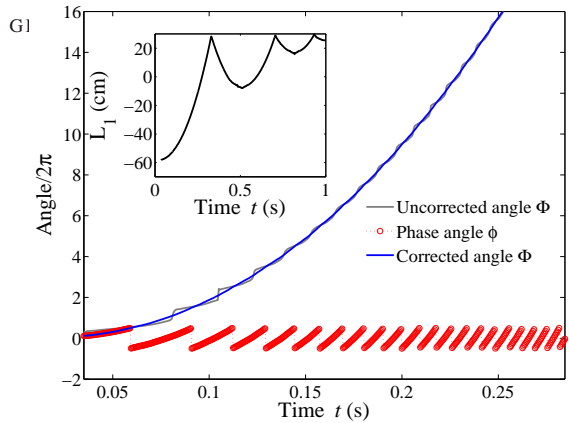}
\caption{Arc-tangential demodulation process to obtain the traveling distance from the Tx. to a Rx. antenna. The red open symbols correspond to the outcome from the demodulation and the blue curve represents the continuous phase shift that scales with the traveling distance of an EM wave. The gray curve corresponds to the $\Phi$ without correcting IQ mismatch. Inset shows an example of the variation of $L_{\rm 1}$ over a longer time.}
\label{fig:dist}
\end{figure}

After the correction of IQ mismatch, the corresponding phase angles are obtained by $\phi = \arctan(Q/I)$. Because $\phi$ is a modulo of $2\pi$, a further correction on the phase jump is needed to obtain the continuous phase shift $\Phi$. In this step, a threshold is introduced to determine whether a phase jump occurs or not and in which direction the jump takes place. As the phase shift of the $i$th channel $\Phi_i \propto L_i$, the variation of $\Phi$ with time (see the blue curve in Fig.\,\ref{fig:dist}) indicates that the target object moves initially slow and accelerates while moving away from the antennae. As demonstrated by a comparison between corrected $\Phi$ and uncorrected $\Phi_{\rm uncorr}$ phase shift, the aforementioned correction method can effectively reduce unrealistic fluctuations in the reconstructed curves. As shown in the inset of Fig.\,\ref{fig:dist}, the distance $L_1$ obtained from Rx. antenna represents exactly what is expected: The object falls freely with a growing velocity and bounces back when reaching the container bottom, suggesting that the coefficient of restitution, which measures the relative rebound over impact velocities, can be determined with the radar system. In comparison to the standard high speed imaging technique \cite{Mueller2016}, the radar tracking technique requires less data collection and processing efforts.

\section{Validation of reconstructed trajectory}

From the measured distances $\vec{L} \equiv (L_1, L_2, L_3)$, the reconstructed trajectory can be obtained with a coordinate transformation

\begin{equation}
\label{eq:reconst}
\left(
\begin{array}{c}
x \\
y \\
z
\end{array}
\right)=\frac{\vec{r}-\vec{L}}{\vec{T}},
\end{equation}

\noindent where the vector $\vec{r}$ is chosen to be $0$ as it contributes only to a constant offset to the reconstructed trajectory, the transformation matrix reads

\begin{equation}
\label{eq:trmat}
\vec{T} \equiv \left(
\begin{array}{ccc}
\sin{\theta_{\rm 1}}\cos{\phi_{\rm 1}} & \sin{\theta_{\rm 1}}\sin{\phi_{\rm 1}} & 1+\cos{\theta_{\rm 1}} \\
\sin{\theta_{\rm 2}}\cos{\phi_{\rm 2}} & \sin{\theta_{\rm 2}}\sin{\phi_{\rm 2}} & 1+\cos{\theta_{\rm 2}} \\
\sin{\theta_{\rm 3}}\cos{\phi_{\rm 3}} & \sin{\theta_{\rm 3}}\sin{\phi_{\rm 3}} & 1+\cos{\theta_{\rm 3}} 
\end{array}\right)
\end{equation} 

\noindent with $\theta_i$ and $\phi_i$ the tilt and azimuth angles of the $i$th antenna, respectively. The transformation matrix is determined from a calibration process using the same tracer particle moving in a known trajectory. A detailed description of the calibration and coordinate transformation processes can be found in \cite{Ott2017}. 

\begin{figure}
\vskip -4em
\centering
\includegraphics[width=0.85\columnwidth]{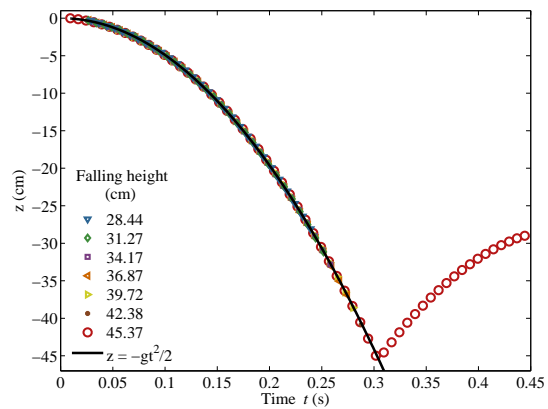}
\caption{A comparison of reconstructed free-falling curves at various initial falling heights. The solid line corresponds to the expected free-falling curve for the largest falling height. Note that the curves for various $H$ are shifted to have the initial falling position $z=0$\,cm, and only the trajectories before the first bouncing with the container bottom are shown except for $H=22.37$\,m. For each curve, one over 15 data points are shown here for a better visibility.}
\label{fig:varHeight}
\end{figure}

Finally, we compare the reconstructed trajectories of a free-falling object from different initial falling heights with the expected parabolic curve. As shown in Fig.\,\ref{fig:varHeight}, the falling curves agree with the expected curve well, demonstrating that, after a proper correction of IQ mismatch, the radar system can be used for particle tracking. Further tests with the Styrofoam container filled with expanded polypropylene (EPP) particles also show that this system can be readily used for the investigation of granular dynamics. Further investigations will focus on particle tracking with various types of granular materials, particularly how to deal with distorted signals arising from the multiple scattering of the surrounding particles. 

\section{CONCLUSION}

To summarize, this investigation suggests that advances in radar tracking technology can be helpful in the investigation of granular dynamics. Using an X-band continuous wave radar system, we are able to track a centimeter sized metallic object in 3D, which enables, for instance, a measurement of the coefficient of restitution of the particle. In comparison to other particle imaging techniques already being used for granular particles \cite{Amon2017}, continuous-wave radar tracking has the advantage of high time resolution and low data collection and processing requirements. With the rapid development of radar technology, this approach is also expected to be more cost effective and accurate. 

Moreover, we show that the accuracy of the radar tracking technique depends strongly on a proper correction of  IQ mismatch, which arises predominately from the mobility of the tracer itself. A practical approach has been proposed to correct the instantaneously changing bias as well as gain errors in the raw IQ signals. Finally, we validate this approach through an analysis on the reconstructed trajectories of a free-falling sphere.

\section*{Acknowledgment}

We acknowledge Felix Ott for his preliminary work on the experimental set-up and Klaus Oetter for technical support. Helpful discussions with Valentin Dichtl, Simeon V\"olkel and Ingo Rehberg are gratefully acknowledged. This work is partly supported by German Research Foundation through Grant No.~HU1939/4-1.

\vspace{6mm}
{\par\noindent}{\Large \bf Bibliographies} 


\aubio{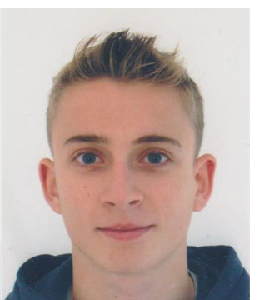}{\textbf{Felix Rech} received his Bachelor of Science degree in physics from the University of Bayreuth in 2018. He is now a master student at the technical University of Darmstadt, aimimg to finish his studies in October 2020.}

\vskip 2em

\aubio{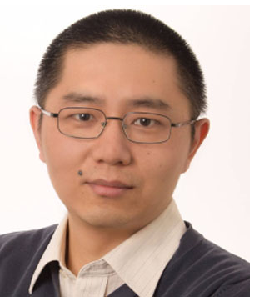}{\textbf{Kai Huang} received his Bachelor of Science degree in Electronic Science and Engineering in 2000 and his Ph.D. in Physical Acoustics, both from Nanjing University in China. He did postdoctoral training (2006-2009) at Max-Planck Institute for Dynamics and Self-organization. Since 2009, he started to build a research group in the Institute of Physics at the University of Bayreuth in Germany. In 2014, he got his 'Habilitation' in physics. From 2015 to 2019, he continued to work at University of Bayreuth as a ‘Privatdozent’. Currently, he is an associate professor in physics at Duke Kunshan University in China. He is interested in using particle tracking techniques to understand, predict and eventually control the collective behavior of granular materials, such as sands, powders and grains. He is also interested in room acoustics, particularly the acoustical design of opera theatres in Europe as well as in China.}

\end{document}